\documentclass[aps,prd,reprint,longbibliography,onecolumn,superscriptaddress,12pt, tightenlines]{revtex4-2}

\usepackage{amsmath,amssymb,mathtools}
\usepackage{bm}
\usepackage{graphicx}

\def\faz{f(x_e)\,\left[1 - (\alpha Z)^2/2\right]}
\newcommand{\diff}{\mathrm{d}}
\newcommand{\UAlberta}{Department of Physics, University of Alberta, Edmonton, Alberta, Canada T6G 2G7}
\newcommand{\Shaanxi}{School of Physics and Information Technology, Shaanxi Normal University, Xi'an 710119, China}

\makeatletter
\renewcommand\normalsize{%
  \@setfontsize\normalsize{12pt}{14pt}%
}
\makeatother

\begin{document}

\title{Total decay rate of a muon bound to a light nucleus}

\author{A.~Czarnecki}
\email{andrzejc@ualberta.ca}
\affiliation{\UAlberta}

\author{A.~O.~Davydov}
\affiliation{\UAlberta}

\author{M.~Y.~Kaygorodov}
\affiliation{\Shaanxi}

\begin{abstract}
We revisit the total decay rate of a muon being in the ground state of a Coulomb potential with atomic charge numbers $4\leq Z \leq 9$.
The discrepancy between the perturbative $(\alpha Z)^2$ result of [Phys.\ Rev.\ \textbf{119}, 365 (1960)] and the fully relativistic partial-wave calculation of [At.\ Data Nucl.\ Data Tables~\textbf{54},~165~(1993)] for oxygen ($Z=8$) is shown to originate from insufficient convergence of the partial-wave series in the latter work.
Our accurate relativistic calculations restore agreement with the perturbative $\alpha Z$~expansion and indicate a negative sign for the next-order $(\alpha Z)^3$ correction.
\end{abstract}

\maketitle

\section{Introduction}
Muonic atoms, in which a negatively charged muon is bound to a nucleus, are used in the determination of weak-interaction parameters~\cite{Kammel:2021jss} and nuclear-structure properties~\cite{Krauth:2021foz}. 
Muonic atoms also offer sensitive probes of physics beyond the Standard Model~\cite{Uesaka:2024tfn,Matias:2025vrq}.
In the near future, several experiments~\cite{Miscetti:2025uxk,COMET:2025sdw,Nishiguchi:2025bsr} will employ muonic atoms to improve sensitivity to the charged-lepton-flavor-violating (CLFV) process of coherent muon-to-electron conversion in the nuclear field, \(\mu^- N \to e^- N\), by up to four orders of magnitude.
\par
The signature of this neutrinoless conversion process is a mono-energetic electron emitted with an energy \(E_e \simeq m_\mu - E_b\), where \(m_\mu\) and \(E_b\) are, respectively, the mass and the binding energy of the muon.
Another branch of the decay which does not violate flavor conservation is the standard muon decay $\mu^- N \longrightarrow e^-N + \bar{\nu}_e + \nu_\mu$, where the initial-state muon is bound to a nuclear potential and the final-state electron is unbound.
This process, also referred to as ``muon decay in orbit'' (DIO), is characterized by a continuous energy distribution of the emitted electron and extends up to the characteristic energy of the electron in the CLFV conversion process.
Consequently, the DIO process constitutes the physical background to the muon-to-electron conversion, and theoretical understanding of the DIO properties directly impacts the sensitivity of these CLFV searches.
\par
The DIO spectrum has been the subject of extensive theoretical study. 
For systems of experimental interest, the influence of the nuclear
recoil~\cite{Czarnecki:2011mx}, leading quantum-electrodynamical
effects~\cite{Szafron:2016cbv}, finite nuclear
size~\cite{Heeck:2021adh}, nuclear deformation and electron screening~\cite{Kaygorodov:2025yag} on the electron spectrum was analyzed.
A refined effective field theory to describe the DIO process was
recently developed in
Ref.~\cite{Fontes:2025xbt,Fontes:2025mps,Fontes:2024yvw}.
\par
The integrated DIO electron spectrum, i.e. the \textit{total} decay rate, provides a valuable global check on the frameworks that underlie theoretical calculations.
For a point-like nucleus, for a massless electron, and for the ground state of the bound muon, the total DIO rate, $\Gamma$, was calculated in Ref.~\cite{Uberall:1960zz} using perturbation theory accurate up to the order~$(\alpha Z)^2$
\begin{equation}
  \frac{\Gamma}{\Gamma_0}
  = 1-\frac{(\alpha Z)^2}{2}+\mathcal{O}(\alpha Z)^3.
  \label{eq:Uberall-smallZ}
\end{equation}
Thus, the total DIO rate differs only slightly from the decay rate of
a free muon,
$\Gamma_0 = G_{\mathrm{F}}^{2}
m_\mu^{5}/(192\pi^{3})$~\cite{MuLan:2010shf}, where $G_{\mathrm{F}}$
is the Fermi constant, and units $\hbar=c=1$ are implied throughout
the paper.  The $(\alpha Z)^2$ term arises from an interplay between
two partially canceled effects.  On the one hand, the muon binding
energy reduces the phase space of the final particles.  On the other
hand, the Coulomb potential enhances the magnitude of the electron
wave function in the nuclear region where the muon wave function is
localized.
\par
In contrast, fully relativistic calculations of the total DIO rate based on a partial-wave expansion~\cite{Watanabe:1993emp} found that for oxygen ($Z=8$) and for the ground state of the muon $1-\Gamma/\Gamma_0 = 0.006$. 
This value is substantially larger, by roughly a factor of three, than $(\alpha Z)^2/2 \simeq 0.002$ obtained from Eq.~\eqref{eq:Uberall-smallZ}.
However, the expected next-to-leading $(\alpha Z)^3$ correction,
$(\alpha Z)^3 \sim 2\times10^{-4}$, is more than an order of magnitude
too small to account for this difference and cannot explain the deviation between the perturbative $\alpha Z$ expansion and the fully-relativistic framework based on the partial-wave expansion.
\par
The inconsistency for $Z=8$ between these two different methods motivates us to reexamine the total DIO rate for light nuclei.
In this work we compute the rate for a muon being in the ground state of the Coulomb potential for atomic numbers $4 \leq Z \leq 9$. 
We treat the nucleus as static and point-like (finite-size effects are negligible compared with the discrepancy mentioned above) and describe the initial-state muon and the final-state electron with Dirac-Coulomb wave functions, following the relativistic approach of Refs.~\cite{Watanabe:1993emp,Watanabe:1987su}.
Nuclear recoil, while relevant near the endpoint of the electron spectrum, does not affect the total decay rate at our target precision~\cite{Czarnecki:2011mx}.
Our primary goals are to determine whether the discrepancy originates from an intrinsic limitation of the $\alpha Z$ expansion, a numerical issue in the earlier partial-wave calculation, or another source, and to establish a reliable total DIO rate benchmark for light muonic atoms.
\par
The paper is organized as follows. Section~\ref{Formalism} introduces the theoretical methods and presents the partial–wave expansion of the total decay rate of a bound muon in the ground state. 
Section~\ref{sec:numerics} contains details of the numerical calculations, shows results for the total DIO rate, and compares them with previous calculations.
Finally, Section~\ref{chap:Conclusion} summarizes our conclusions.

\section{Theoretical method}\label{Formalism}

To obtain an expression for the muon decay rate, we employ the formalism presented in Ref.~\cite{Aslam:2020pqn}. 
Treating the nucleus as a source of a classical electromagnetic field, the three-body decay $(Z\mu^-)  \to [Ze^-]\,\bar{\nu}_e\,\nu_\mu$ can be represented as two sequential two-body decays,
$(Z\mu^-) \to [Ze^-]A$ and $A \to \bar{\nu}_e \nu_\mu$, where $A$ is a fictitious neutral massive spin-1 boson carrying the lepton–flavor quantum numbers of the neutrino pair.
The notation $(Z\mu^-)$ denotes that the muon is bound to the nuclear
potential and $[Ze^-]$ means that the electron is unbound in the field of the nuclear potential.
The factorization of the $(Z\mu^{-}) \to [Ze^{-}]\,\bar{\nu}_e\,\nu_\mu$  total decay rate averaged over the initial states of the muon and summed over the final states of the electron, neutrino, and antineutrino can be written as
\begin{equation}
\label{eq:intermediate-boson-mA}
\Gamma\!\left((Z\mu^-) \to [Ze^-]\,\bar{\nu}_e \nu_\mu\right)
=
\frac{12288\,\pi^{2}\Gamma_0}{g^{4} m_\mu^{5}}
\int_{0}^{m_{A,\max}}\! \diff m_A\;
m_A^{2}\,
\Gamma\!\left((Z\mu^-)\to [Ze^-]A\right)\,
\Gamma\!\left(A\to \bar{\nu}_e\nu_\mu\right),
\end{equation}
where $g$ is the weak-coupling constant, $m_A$ is the invariant mass of the boson~$A$, $m_{A,\max}=E_\mu-E_e$, with $E_\mu$ and $E_e$ being the total energies of the initial-state muon and the final-state electron, respectively.
An advantage of this factorization is that all atomic effects (and hence the $\alpha Z $ dependence of $\Gamma((Z\mu^{-}) \to [Ze^-]\,\bar{\nu}_e\,\nu_\mu)$) are encoded in the total two-body decay rate
$\Gamma((Z\mu^{-})\to [Ze^-]A)$, whereas $\Gamma(A\to \bar{\nu}_e\nu_\mu)$ is universal and does not depend on properties of the nuclear field.
The total decay rate $\Gamma(A\to \bar{\nu}_e\nu_\mu)$ can be evaluated analytically in the limit of massless neutrinos, see e.g. Ref.~\cite{Aslam:2020pqn}.
Substituting the expression for $\Gamma(A\to \bar{\nu}_e\nu_\mu)$ and introducing
\begin{equation}
  z \equiv \frac{m_A}{m_\mu}, \qquad
  0 \le z \le z_{\max} = \frac{E_\mu-E_e}{m_\mu}\,,
\end{equation}
we obtain~\cite{Aslam:2020pqn}
\begin{equation}
\label{eq:intermediate-boson}
\Gamma\!\left((Z\mu^-) \to [Ze^-]\,\bar{\nu}_e \nu_\mu\right)
=\frac{256\pi\,\Gamma_0}{g^{2} m_\mu}\int_{0}^{z_{\max}}\! \diff z\;
\Gamma\!\left((Z\mu^-) \to [Ze^-]A\right)\, z^{3}.
\end{equation}
Equation~\eqref{eq:intermediate-boson} is valid for the bound-state muon as well as for the free muon. 
The two-body total decay rate $\Gamma((Z\mu^-)\to [Ze^-]A)$ of the muon decay into the electron and the fictitious boson $A$ can be written as
\begin{equation}
\label{eq:Gamma-intermediate}
\Gamma\!\left((Z\mu^-) \to [Ze^-]A\right)
=\frac{1}{64\pi^5}\!\int\!\frac{\diff^{3}\vec{p}_e}{p_eE_e}\;
      \frac{\diff^{3}\vec{k}}{k_{0}}\;
      \delta\!\left(E_\mu-E_e-k_{0}\right)
      \bigl|{\cal M}_{(Z\mu^-)\to [Ze^-]A}\bigr|^{2},
\end{equation}
where $\bigl|{\cal M}_{(Z\mu^-)\to [Ze^-]A}\bigr|^{2}$ is the square of the absolute value of the amplitude of the process averaged over the initial states of the muon and summed over the final states of the electron and the fictitious boson $A$, $p_e^\nu=(E_e,\vec{p}_e)$ with $p_e\equiv|\vec{p}_e|$ and 
$k^\nu=(k_0,\vec{k})$ are the four-momenta of the final-state electron and the boson $A$, respectively.

Taking into account that the nuclear potential is spherically symmetric, it is natural to evaluate $\Gamma((Z\mu^-)\to [Ze^-]A)$ entering Eq.~\eqref{eq:Gamma-intermediate} using a relativistic central-field approximation for the muon and electron.
The ground-state muon wave function is given by
\begin{equation}\label{muon wf}
\Psi_{\mu_\mu}(\vec r)
=\frac{1}{r}
\begin{pmatrix}
G(r)\,\Omega_{-1\mu_\mu}(\hat r)\\[4pt]
i\,F(r)\,\Omega_{1\mu_\mu}(\hat r)
\end{pmatrix},
\end{equation}
where $G$ and $F$ are, respectively, the large and small radial components of the muon wave function, $\Omega_{\kappa\mu}$ is the spherical spinor~\cite{Varshalovich:1988ifq},  $\hat r \equiv \vec r/r$, $\kappa$ is the relativistic angular quantum number ($\kappa = -1$ for the ground state), and $\mu$ is the projection of the total angular momentum. 
The muon radial wave functions are normalized to unity,
\begin{equation}
\int_0^\infty
\bigl[
  G^{2}(r)+F^{2}(r)
\bigr]\,dr
= 1.
\end{equation}
The wave function of the final-state electron is given by
\begin{equation}\label{electron wf}
\psi_{E_e\kappa_e\mu_e}(\vec r)
=\frac{1}{r}
\begin{pmatrix}
g_{E_e\kappa_e}(r)\,\Omega_{\kappa_e\mu_e}(\hat r)\\[4pt]
i\,f_{E_e\kappa_e}(r)\,\Omega_{-\kappa_e\mu_e}(\hat r)
\end{pmatrix},
\end{equation}
$g_{E_e\kappa_e}$ and $f_{E_e\kappa_e}$ are, respectively, the large and small radial components of the electron wave function normalized in the energy scale,
\begin{equation}
\int_0^\infty
\bigl[
  g_{E_e\kappa_e}(r)\,g_{E'_e\kappa_e}(r)
 +f_{E_e\kappa_e}(r)\,f_{E'_e\kappa_e}(r)
\bigr]\,dr
= \delta(E_e-E'_e).
\end{equation}
Introducing the leptonic current,
\begin{equation}
\label{eq:current muon-electron}
J_{\alpha}(\vec{q})
=\int \diff^{3}\vec{r}\,
e^{-i\vec{q}\cdot\vec{r}}\,
\psi^{\dagger}_{E_e\kappa_e\mu_e}(\vec{r})\,\gamma_{0}\gamma_{\alpha}P_{L}\,
\Psi_{\mu_\mu}(\vec{r}),
\end{equation}
where $P_{L}=(1-\gamma_{5})/2$, the averaged-summed absolute value squared of the amplitude $\bigl|{\cal M}_{(Z\mu^-)\to [Ze^-]A}\bigr|^{2}$ can be written as
\begin{equation}
\label{current}
\bigl|{\cal M}_{(Z\mu^-)\to [Ze^-]A}\bigr|^{2}
=
\frac{1}{2}
\sum_{\mu_\mu}\sum_{\kappa_e\mu_e}
\frac{g^{2}}{2}\,
J^{\alpha}(\vec{k})\,
J^{\dagger\beta}(\vec{k})\,
\left(-g_{\alpha\beta}+\frac{k_{\alpha}k_{\beta}}{m_A^{2}}\right),
\end{equation}
where the tensor in parentheses stems from the polarization sum
for a massive spin-1 boson with four-momentum $k$ and
$g_{\alpha\beta}=\mathrm{diag}(1,-1,-1,-1)$ is the Minkowski metric tensor.

To obtain the final expression for $\Gamma\!\left((Z\mu^-) \to [Ze^-]\,\bar{\nu}_e \nu_\mu\right)$, we represent the effective leptonic current~$J^{\alpha}$ using the multipole expansion for~$e^{i\vec{k}\cdot\vec{r}}$~\cite{Varshalovich:1988ifq}, sum over the total angular momentum projections $\mu_{\mu}$ and $\mu_e$, contract the Lorentz indices in Eq.~(\ref{current}), and finally integrate over the angle of~$\vec{k}$ in Eq.~(\ref{eq:Gamma-intermediate}). These steps bring us to
the partial-wave expansion of the total decay rate for the ground state of the muon:
\begin{equation}\label{Gamma}
\begin{aligned}
\dfrac{\Gamma}{\Gamma_{0}} 
&
=\sum_{L\kappa_e}\dfrac{8}{m_{\mu}^{5}}\left(2j_{e}+1\right)\!\int_{0}^{E_{\mu}}\!\diff
E_{e}\!\int_{0}^{E_{\mu}-E_{e}}\!k^{2}\diff k \\
 & \times
\Biggl\{\left[\left(E_{\mu}-E_{e}\right)^{2}-k^{2}\right]
\!\left(\dfrac{\vert S_{\kappa_e L}^{0}\vert^{2}}{L\left(L+1\right)}+\dfrac{\vert S_{\kappa_e L}^{-1}\vert^{2}}{L\left(2L+1\right)}+\dfrac{\vert S_{\kappa_e L}^{+1}\vert^{2}}{\left(L+1\right)\left(2L+1\right)}\right)\\
 & +k^{2}\!\left(\vert S_{\kappa_e L}\vert^{2}+\dfrac{\vert S_{\kappa_e L}^{+1}+S_{\kappa_e L}^{-1}\vert^{2}}{\left(2L+1\right)^{2}}\right)
 +2\left(E_{\mu}-E_{e}\right)k\mathrm{Im}\!\left[\dfrac{S_{\kappa_e L}\!\left(S_{\kappa_e L}^{-1*}+S_{\kappa_e L}^{+1*}\right)}{2L+1}\right]\!\Biggr\}
 \equiv \sum_{\kappa_e}\frac{\Gamma_{\kappa_e}}{\Gamma_0}\,
\end{aligned}
\end{equation}
where $\mathrm{Im}(\cdots)$ denotes the imaginary part and
$j_e = |\kappa_e| - \tfrac{1}{2}$ is the total angular momentum of the electron.
The radial integrals $S^{0}_{\kappa_e L}$, $S^{-1}_{\kappa_e L}$,
$S^{+1}_{\kappa_e L}$, and $S_{\kappa_e L}$ are given by
\begin{equation}
\label{S-functions}
\begin{aligned}
S^{0}_{\kappa_e L} &=
\begin{cases}
-i(\kappa_e-1)\,\langle j_{L}(kr)\,(f_{E_e\kappa_e}G+g_{E_e\kappa_e}F)\rangle, & \kappa_e<0,\\
\phantom{-}(\kappa_e+1)\,\langle j_{L}(kr)\,(g_{E_e\kappa_e}G-f_{E_e\kappa_e}F)\rangle, & \kappa_e>0,
\end{cases}\\[4pt]
S^{-1}_{\kappa_e L} &=
\begin{cases}
\langle j_{L-1}(kr)\,[(\kappa_e-L-1)g_{E_e\kappa_e}G-(\kappa_e+L-1)f_{E_e\kappa_e}F]\rangle \\
\qquad -\,i\,\langle j_{L-1}(kr)\,[\,(\kappa_e+L+1)f_{E_e\kappa_e}G+(\kappa_e-L+1)g_{E_e\kappa_e}F\,]\rangle, & \kappa_e<0,\\
\text{(analogous with }\kappa_e\!\to\!-\kappa_e\text{)}, & \kappa_e>0,
\end{cases}\\[4pt]
S^{+1}_{\kappa_e L} &=
\begin{cases}
\langle j_{L+1}(kr)\,[\,(\kappa_e+L)g_{E_e\kappa_e}G+(-\kappa_e+L+2)f_{E_e\kappa_e}F\,]\rangle \\
\qquad -\,i\,\langle j_{L+1}(kr)\,[\,(\kappa_e-L)f_{E_e\kappa_e}G+(\kappa_e+L+2)g_{E_e\kappa_e}F\,]\rangle, & \kappa_e<0,\\
\text{(analogous with }\kappa_e\!\to\!-\kappa_e\text{)}, & \kappa_e>0,
\end{cases}\\[4pt]
S_{\kappa_e L} &=
\begin{cases}
i\,\langle j_{L}(kr)\,(f_{E_e\kappa_e}G-g_{E_e\kappa_e}F)\rangle, & \kappa_e<0,\\
\phantom{i}\,\langle j_{L}(kr)\,(g_{E_e\kappa_e}G+f_{E_e\kappa_e}F)\rangle, & \kappa_e>0,
\end{cases}
\end{aligned}
\end{equation}
where we used the following notation: $\langle \cdots \rangle\equiv\int_{0}^{\infty}\!\cdots\,\,\diff r$.
Here, $L$ labels the multipole terms in the partial-wave expansion of the leptonic current $J^\alpha$, and $j_L(x)$ are spherical Bessel functions.
Selection rules imply $\kappa_e=\pm L$ or $\pm(L+1)$ (with $L\neq 0$ contributing to $S_{\kappa_e L}$ and $S^{0}_{\kappa_e L}$).
A detailed derivation of the partial–wave expansion of the bound-muon decay rate for arbitrary initial state of the muon is presented in Ref.~\cite{Kaygorodov:2025yag}.

\section{Numerical results and discussion}\label{sec:numerics}
Using Eqs.~\eqref{Gamma} and \eqref{S-functions}, we perform fully relativistic calculations of the normalized total decay rate, $\Gamma/\Gamma_0$, for the muon bound to the point-like Coulomb potential, for atomic charge numbers in the range $4 \leq Z \leq 9$. 
The bound muon is described with the analytical Dirac-Coulomb wave function~\cite{Rose:1961RET}.
The wave function of the outgoing unbound electron is obtained by solving the Dirac equation numerically using the \textsc{RADIAL} package~\cite{salvat1995radial,salvat2019radial} in double-precision arithmetic. 
The oscillatory radial integrals that appear in the partial-wave expansion of the total decay rate,~\eqref{S-functions}, are evaluated using the adaptive Gauss-Kronrod routine \texttt{DQAGS} from the library \textsc{QUADPACK}~\cite{piessens2012quadpack}, with both absolute and relative error tolerances set to $10^{-9}$.
\par
The main numerical challenge in this work is the convergence of the partial-wave expansion of the total decay rate. 
For each value of $Z$, we sum partial-wave contributions~$\Gamma_{\kappa_e}/\Gamma_0$ until the magnitude of successive terms falls below a threshold $10^{-5}$-$10^{-6}$. 
In order to estimate the numerical uncertainty associated with the
terminated partial-wave expansion, we approximated the last terms of the partial-wave series with a geometric series and summed the extrapolated tail analytically. 
This conservative extrapolation procedure yields a residual contribution to $\Gamma/\Gamma_0$ about $10^{-5}$-$10^{-6}$ for all values of $Z$ considered, which we take as an upper bound on the numerical uncertainty associated with termination of the partial-wave expansion.

\begin{table}[htb]
\caption{The total bound-muon decay rate normalized to that of the free-muon decay rate, $\Gamma/\Gamma_{0}$, for various values of atomic charge number $Z$.
The second and third columns show results of the fully-relativistic calculation of the present work (PW) and Ref.~\cite{Watanabe:1993emp}, respectively.
The fourth column shows results obtained using the perturbative $\alpha Z$ expansion of Ref.~\cite{Uberall:1960zz} corrected to account for the finite electron mass. 
The fifth column,~$\Delta(Z)$, shows the difference between the $\alpha Z$ results and the PW ones in units of $10^{-3}$.}
\label{tab:ratio}
\begin{ruledtabular}
\begin{tabular}{c c c c c}
  $Z$
  & PW
  & Ref.~\cite{Watanabe:1993emp}
  & Ref.~\cite{Uberall:1960zz}
  & $\Delta(Z)\,[10^{-3}]$
   \\
\hline
 4 & 0.9993 &       & 0.9994 & 0.045        \\
 5 & 0.9990 &       & 0.9991 & 0.099        \\
 6 & 0.9987 &       & 0.9989 & 0.188        \\
 7 & 0.9982 &       & 0.9985 & 0.322        \\
 8 & 0.9976 & 0.994 & 0.9981 & 0.506        \\
 9 & 0.9969 &       & 0.9977 & 0.757        \\
\end{tabular}
\end{ruledtabular}
\end{table}
In Table~\ref{tab:ratio} we present the converged results for the normalized bound-muon total decay rate, $\Gamma/\Gamma_0$, and compare them with the results of Refs.~\cite{Uberall:1960zz,Watanabe:1993emp}. 
In Ref.~\cite{Watanabe:1993emp} the fully-relativistic calculations
based on the partial-wave expansion were performed.
In Ref.~\cite{Uberall:1960zz}, the semi-relativistic wave functions accurate up to the order~$(\alpha Z)^2$ were used for both the muon and the electron.
For a consistent comparison with the analytical result
Eq.~(\ref{eq:Uberall-smallZ}) of Ref.~\cite{Uberall:1960zz},  derived for a massless electron, we apply a phase-space correction factor~\cite{Aslam:2020pqn},
\begin{equation}\label{eq:fxe}
f(x_e) = 1 - 8x_e^2 - 24x_e^4\ln x_e + 8x_e^6 - x_e^8 = 0.99981, \quad
x_e \equiv \frac{m_e}{m_\mu} = 4.8363 \times 10^{-3},
\end{equation}
to account for the finite electron mass. 
This factor exactly reproduces the electron–mass dependence of the free-muon decay rate~\cite{Aslam:2020pqn}, and when used as a multiplicative correction to Ref.~\cite{Uberall:1960zz} bound-muon result it captures the leading mass effect with an error below the numerical precision of our calculation.
This correction changes the total decay rate of Ref.~\cite{Uberall:1960zz} by approximately $0.01 \%$.
The difference
\begin{equation}
  \Delta(Z) = f(x_e)\bigl[1 - (\alpha Z)^2/2\bigr] - \Gamma/\Gamma_0,
\end{equation}
shown in the last column of Table~\ref{tab:ratio}, quantifies the contribution from higher-order terms in the $\alpha Z$ expansion beyond $(\alpha Z)^2$. 
For oxygen ($Z=8$), the present value $1 - \Gamma/\Gamma_0 = 0.0024$ is notably closer to the corrected result of Ref.~\cite{Uberall:1960zz}, $1-f(x_e)\,\left[1-(\alpha Z)^2/2\right] \simeq 0.0020$, than to the value $1-\Gamma/\Gamma_0=0.006$ reported in Ref.~\cite{Watanabe:1993emp}.

\begin{figure}[htb]
\centering
\includegraphics[scale=1.0]{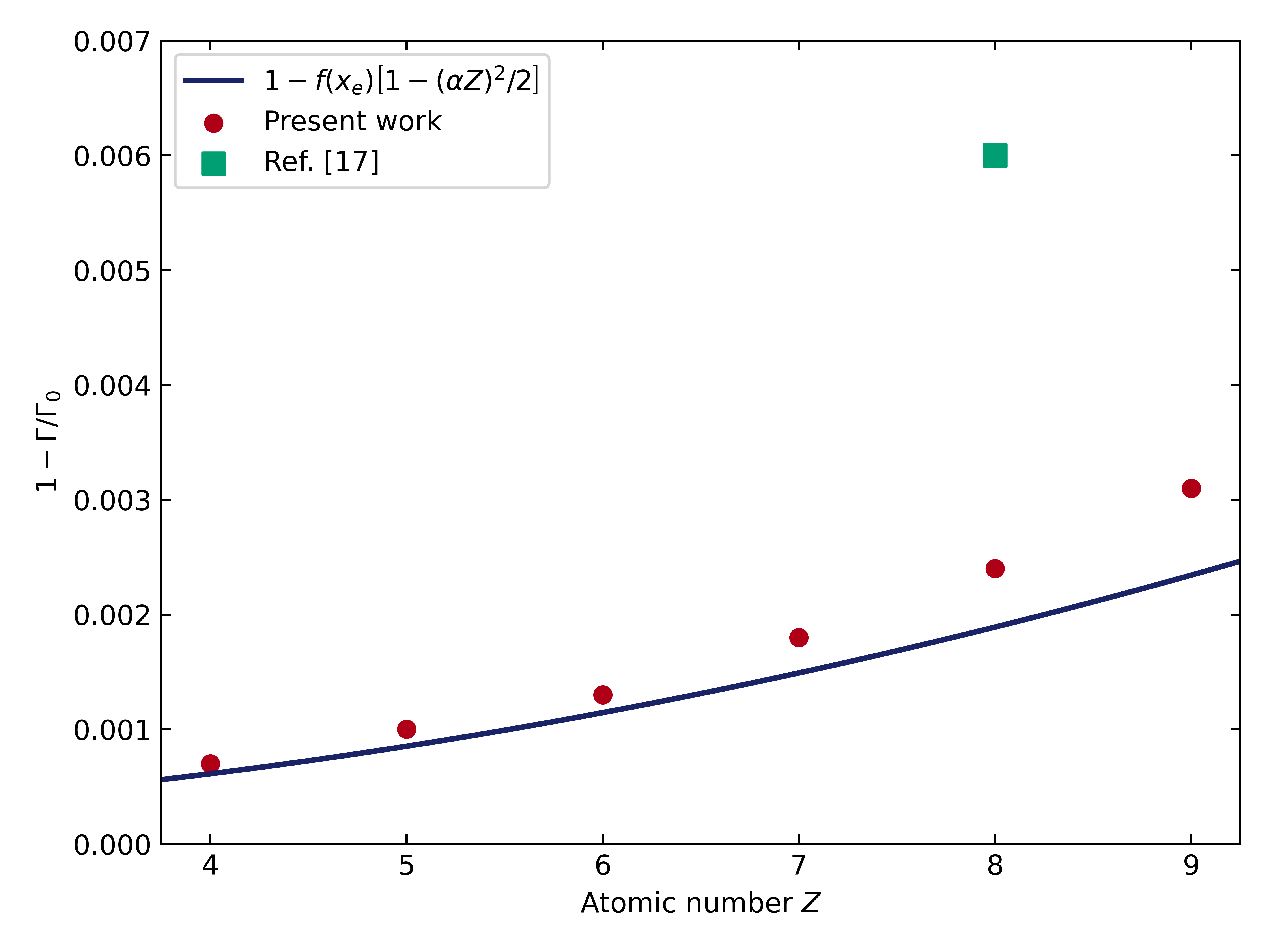} 
\caption{Comparison of the present converged values for $1-\Gamma/\Gamma_0$ (red circles) with $1-\faz$ (blue solid line), where $f(x_e)$ is a correction due to the finite electron mass, and
the result of Ref.~\cite{Watanabe:1993emp} (green square).
The difference between the red circles and the solid line is interpreted as the next-order-in-$\alpha Z$ corrections.
}
\label{fig:uberall}
\end{figure}

The systematic behavior of our results is visualized in Fig.~\ref{fig:uberall}, which plots $1 - \Gamma/\Gamma_0$ versus the atomic number $Z$. 
The solid line represents the corrected $(\alpha Z)^2$ prediction $1 - f(x_e)\left[1-(\alpha Z)^2/2\right]$ of Ref.~\cite{Uberall:1960zz}, while the red circles denote the present all-order-in-$\alpha Z$ values. 
Our relativistic results for $1 - \Gamma/\Gamma_0$ lie consistently above the $(\alpha Z)^2$ curve. 
The magnitude of this deviation grows with $Z$, as quantified by the positive $\Delta(Z)$ values in Table~\ref{tab:ratio}. 
This systematic trend is consistent with a small negative next-order-in-$\alpha Z$ correction to $\Gamma/\Gamma_0$ in the $\alpha Z$ expansion of the bound-muon decay rate. 
Consequently, the next-order term in the $\alpha Z$ expansion for light nuclei must be \textit{negative}. 
In contrast, the result of Ref.~\cite{Watanabe:1993emp} for $Z=8$ (green square in Fig.~\ref{fig:uberall}) significantly deviates from both our data and the $(\alpha Z)^2$ trend.

To identify the source of the discrepancy noted above, we examine in detail the convergence of the partial-wave expansion for $Z=8$.
\begin{table}[htb] 
\caption{Convergence of the partial-wave expansion for ${}^{16}\mathrm{O}$ ($Z=8$). 
The table shows the partial sum $\sum_{|\kappa_e|\le\kappa_{\max}}\Gamma_{\kappa_e}/\Gamma_0$ as a function of the cutoff $\kappa_{\max}$.}
\label{Z=8}
\begin{ruledtabular}
\begin{tabular}{c c c}
$\kappa_{\max}$ & Present Work & Ref.~\cite{Watanabe:1993emp} \\
  \hline
25 & 0.98871 & \\
29 & 0.99389 & 0.994 \\
35 & 0.99661 & \\
40 & 0.99728 & \\
45 & 0.99750 & \\
50 & 0.99757 & \\
55 & 0.99759 & \\
59 & 0.99760 & \\
\end{tabular}
\end{ruledtabular}
\end{table}
Table~\ref{Z=8} presents the partial sums
$\sum_{|\kappa_e|\le\kappa_{\max}}\Gamma_{\kappa_e}/\Gamma_0$ as a
function of the parameter $\kappa_{\max}$ at which the partial-wave
series is terminated.  
In Ref.~\cite{Watanabe:1993emp}, the partial wave expansion of the total decay rate was truncated at $L=31$, which results in $\Gamma/\Gamma_0 = 0.994$.
This value is numerically very close to our partial sum with
$\kappa_{\mathrm{max}} =29$, $\Gamma/\Gamma_0 = 0.99389$. 
Our converged result, obtained by extending the sum to $\kappa_{\mathrm{max}} = 59$ and including the extrapolation of the remaining tail, is $\Gamma/\Gamma_0 = 0.99760$. 
Thus, the result of Ref.~\cite{Watanabe:1993emp} 
corresponds to a partial-wave expansion that was terminated before the series had converged.
The discrepancy between the corrected $(\alpha Z)^2$ result of Ref.~\cite{Uberall:1960zz} and the fully relativistic calculation of Ref.~\cite{Watanabe:1993emp} is therefore resolved once the partial-wave expansion is carried up to sufficient $\kappa_{\max}$.

\section{Conclusion}\label{chap:Conclusion}
In this work we have reexamined the total decay rate of a muon bound to a point-like Coulomb potential for atomic charge numbers $4\leq Z \leq 9$. 
To this end, we employed a fully relativistic formalism based on the partial wave expansion of the bound-muon total decay rate and numerical solution of the Dirac equation in the Coulomb potential. 
By carefully monitoring the convergence of the partial-wave series, we find that our numerical results are consistent with the analytical result of Ref.~\cite{Uberall:1960zz}, which is accurate to the order $(\alpha Z)^2$.
\par
Our all-order-in-$\alpha Z$ relativistic calculations yield values of $\Gamma/\Gamma_0$ that are systematically smaller than the corresponding $(\alpha Z)^2$ prediction of Ref.~\cite{Uberall:1960zz} corrected to account for the finite electron mass~\cite{Aslam:2020pqn}. 
Thus, the present calculations provide numerical evidence for a negative next-order-in-$\alpha Z$ correction to $\Gamma/\Gamma_0$ for light nuclei.
\par
For $Z=8$, we identified a significant disagreement between the fully relativistic result of Ref.~\cite{Watanabe:1993emp} and both the analytical $(\alpha Z)^2$ prediction of Ref.~\cite{Uberall:1960zz} and our numerical relativistic results. 
We showed that this discrepancy is due to an insufficiently converged partial-wave expansion in the calculation of Ref.~\cite{Watanabe:1993emp}.
Specifically, the termination of the partial-wave series at $\kappa_{\max} = 29$ used in Ref.~\cite{Watanabe:1993emp} resulted in $\Gamma / \Gamma_0 = 0.994$, whereas the analytical result of Ref.~\cite{Uberall:1960zz} gives $0.9981$. 
The present work shows that termination of the partial-wave expansion at $\kappa_{\max}=59$ with a subsequent extrapolation of the tail yields  $\Gamma / \Gamma_0 = 0.9976$, restoring the agreement between the perturbative $\alpha Z$ result of Ref.~\cite{Uberall:1960zz} and the fully-relativistic calculations of Ref.~\cite{Watanabe:1993emp}.
\par
In the present calculations of the bound-muon total decay rate we considered only the effect of the Coulomb potential. 
The finite-nuclear-size, electron screening, and quantum electrodynamic effects were not taken into account.
However, for small values of $Z$ the contribution of the above-mentioned corrections to the total decay rate is expected to be smaller than the present numerical accuracy.
For light muonic atoms these corrections mainly change the shape of the electron spectrum close to the endpoint region where the spectrum shape rapidly decreases, see, e.g., Refs.~\cite{Czarnecki:2011mx,Szafron:2016cbv, Kaygorodov:2025yag}.
Therefore, these corrections to the total decay rate are not essential for the present analysis.
The updated values of the bound-muon total decay rates for $Z=4$–$9$ provide
a set of benchmark results for future theoretical investigations of light muonic atoms
and may be used for consistency checks in related bound-muon decay calculations.

\begin{acknowledgments}
This work was supported by the Natural Sciences and Engineering
Research Council of Canada (NSERC).
\end{acknowledgments}


\begin{thebibliography}{24}%
\makeatletter
\providecommand \@ifxundefined [1]{%
 \@ifx{#1\undefined}
}%
\providecommand \@ifnum [1]{%
 \ifnum #1\expandafter \@firstoftwo
 \else \expandafter \@secondoftwo
 \fi
}%
\providecommand \@ifx [1]{%
 \ifx #1\expandafter \@firstoftwo
 \else \expandafter \@secondoftwo
 \fi
}%
\providecommand \natexlab [1]{#1}%
\providecommand \enquote  [1]{``#1''}%
\providecommand \bibnamefont  [1]{#1}%
\providecommand \bibfnamefont [1]{#1}%
\providecommand \citenamefont [1]{#1}%
\providecommand \href@noop [0]{\@secondoftwo}%
\providecommand \href [0]{\begingroup \@sanitize@url \@href}%
\providecommand \@href[1]{\@@startlink{#1}\@@href}%
\providecommand \@@href[1]{\endgroup#1\@@endlink}%
\providecommand \@sanitize@url [0]{\catcode `\\12\catcode `\$12\catcode
  `\&12\catcode `\#12\catcode `\^12\catcode `\_12\catcode `\%12\relax}%
\providecommand \@@startlink[1]{}%
\providecommand \@@endlink[0]{}%
\providecommand \url  [0]{\begingroup\@sanitize@url \@url }%
\providecommand \@url [1]{\endgroup\@href {#1}{\urlprefix }}%
\providecommand \urlprefix  [0]{URL }%
\providecommand \Eprint [0]{\href }%
\providecommand \doibase [0]{https://doi.org/}%
\providecommand \selectlanguage [0]{\@gobble}%
\providecommand \bibinfo  [0]{\@secondoftwo}%
\providecommand \bibfield  [0]{\@secondoftwo}%
\providecommand \translation [1]{[#1]}%
\providecommand \BibitemOpen [0]{}%
\providecommand \bibitemStop [0]{}%
\providecommand \bibitemNoStop [0]{.\EOS\space}%
\providecommand \EOS [0]{\spacefactor3000\relax}%
\providecommand \BibitemShut  [1]{\csname bibitem#1\endcsname}%
\let\auto@bib@innerbib\@empty
\bibitem [{\citenamefont {Kammel}(2021)}]{Kammel:2021jss}%
  \BibitemOpen
  \bibfield  {author} {\bibinfo {author} {\bibfnamefont {P.}~\bibnamefont
  {Kammel}} (\bibinfo {collaboration} {MuSun}),\ }\bibfield  {title} {\bibinfo
  {title} {{MuSun - Muon Capture on the Deuteron}},\ }\href
  {https://doi.org/10.21468/SciPostPhysProc.5.018} {\bibfield  {journal}
  {\bibinfo  {journal} {SciPost Phys. Proc.}\ }\textbf {\bibinfo {volume}
  {5}},\ \bibinfo {pages} {018} (\bibinfo {year} {2021})}\BibitemShut {NoStop}%
\bibitem [{\citenamefont {Krauth}\ \emph {et~al.}(2021)\citenamefont {Krauth}
  \emph {et~al.}}]{Krauth:2021foz}%
  \BibitemOpen
  \bibfield  {author} {\bibinfo {author} {\bibfnamefont {J.~J.}\ \bibnamefont
  {Krauth}} \emph {et~al.},\ }\bibfield  {title} {\bibinfo {title} {{Measuring
  the \ensuremath{\alpha}-particle charge radius with muonic helium-4 ions}},\
  }\href {https://doi.org/10.1038/s41586-021-03183-1} {\bibfield  {journal}
  {\bibinfo  {journal} {Nature}\ }\textbf {\bibinfo {volume} {589}},\ \bibinfo
  {pages} {527} (\bibinfo {year} {2021})}\BibitemShut {NoStop}%
\bibitem [{\citenamefont {Uesaka}\ \emph {et~al.}(2025)\citenamefont {Uesaka},
  \citenamefont {Yamanaka},\ and\ \citenamefont {Kuno}}]{Uesaka:2024tfn}%
  \BibitemOpen
  \bibfield  {author} {\bibinfo {author} {\bibfnamefont {Y.}~\bibnamefont
  {Uesaka}}, \bibinfo {author} {\bibfnamefont {M.}~\bibnamefont {Yamanaka}},\
  and\ \bibinfo {author} {\bibfnamefont {Y.}~\bibnamefont {Kuno}},\ }\bibfield
  {title} {\bibinfo {title} {{$\mu^- \to e^-\gamma$ in a muonic atom as a probe
  for effective lepton flavor-violating operators involving photon fields}},\
  }\href {https://doi.org/10.1103/PhysRevD.111.035017} {\bibfield  {journal}
  {\bibinfo  {journal} {Phys. Rev. D}\ }\textbf {\bibinfo {volume} {111}},\
  \bibinfo {pages} {035017} (\bibinfo {year} {2025})},\ \Eprint
  {https://arxiv.org/abs/2411.10304} {arXiv:2411.10304 [hep-ph]} \BibitemShut
  {NoStop}%
\bibitem [{\citenamefont {Matias}\ \emph {et~al.}(2025)\citenamefont {Matias},
  \citenamefont {Lemos},\ and\ \citenamefont {Dahia}}]{Matias:2025vrq}%
  \BibitemOpen
  \bibfield  {author} {\bibinfo {author} {\bibfnamefont {J.~E.~J.}\
  \bibnamefont {Matias}}, \bibinfo {author} {\bibfnamefont {A.~S.}\
  \bibnamefont {Lemos}},\ and\ \bibinfo {author} {\bibfnamefont
  {F.}~\bibnamefont {Dahia}},\ }\bibfield  {title} {\bibinfo {title} {{Probing
  Short-Distance Modifications of Gravity via Spin-Independent and
  Spin-Dependent Effects in Muonic Atoms}}} (\bibinfo {year} {2025}),\ \bibinfo
  {note} {arxiv:2511.00719}\BibitemShut {NoStop}%
\bibitem [{\citenamefont {Miscetti}(2025)}]{Miscetti:2025uxk}%
  \BibitemOpen
  \bibfield  {author} {\bibinfo {author} {\bibfnamefont {S.}~\bibnamefont
  {Miscetti}} (\bibinfo {collaboration} {Mu2e}),\ }\bibfield  {title} {\bibinfo
  {title} {{Status of the Mu2e experiment}},\ }\href
  {https://doi.org/10.1016/j.nima.2025.170257} {\bibfield  {journal} {\bibinfo
  {journal} {Nucl. Instrum. Meth. A}\ }\textbf {\bibinfo {volume} {1073}},\
  \bibinfo {pages} {170257} (\bibinfo {year} {2025})}\BibitemShut {NoStop}%
\bibitem [{\citenamefont {Aoki}\ \emph {et~al.}(2025)\citenamefont {Aoki} \emph
  {et~al.}}]{COMET:2025sdw}%
  \BibitemOpen
  \bibfield  {author} {\bibinfo {author} {\bibfnamefont {M.}~\bibnamefont
  {Aoki}} \emph {et~al.} (\bibinfo {collaboration} {COMET, MEG, Mu2e, Mu3e}),\
  }\bibfield  {title} {\bibinfo {title} {{Charged Lepton Flavour Violations
  searches with muons: present and future}}} (\bibinfo {year} {2025}),\
  \bibinfo {note} {arxiv:2503.22461}\BibitemShut {NoStop}%
\bibitem [{\citenamefont {Nishiguchi}(2025)}]{Nishiguchi:2025bsr}%
  \BibitemOpen
  \bibfield  {author} {\bibinfo {author} {\bibfnamefont {H.}~\bibnamefont
  {Nishiguchi}},\ }\bibfield  {title} {\bibinfo {title} {{A search for
  muon-to-electron conversion at J-PARC : The COMET experiment}},\ }\href
  {https://doi.org/10.22323/1.476.0469} {\bibfield  {journal} {\bibinfo
  {journal} {PoS}\ }\textbf {\bibinfo {volume} {ICHEP2024}},\ \bibinfo {pages}
  {469} (\bibinfo {year} {2025})}\BibitemShut {NoStop}%
\bibitem [{\citenamefont {Czarnecki}\ \emph {et~al.}(2011)\citenamefont
  {Czarnecki}, \citenamefont {Garcia~i Tormo},\ and\ \citenamefont
  {Marciano}}]{Czarnecki:2011mx}%
  \BibitemOpen
  \bibfield  {author} {\bibinfo {author} {\bibfnamefont {A.}~\bibnamefont
  {Czarnecki}}, \bibinfo {author} {\bibfnamefont {X.}~\bibnamefont {Garcia~i
  Tormo}},\ and\ \bibinfo {author} {\bibfnamefont {W.~J.}\ \bibnamefont
  {Marciano}},\ }\bibfield  {title} {\bibinfo {title} {{Muon decay in orbit:
  spectrum of high-energy electrons}},\ }\href@noop {} {\bibfield  {journal}
  {\bibinfo  {journal} {Phys. Rev.}\ }\textbf {\bibinfo {volume} {D84}},\
  \bibinfo {pages} {013006} (\bibinfo {year} {2011})},\ \Eprint
  {https://arxiv.org/abs/1106.4756} {arXiv:1106.4756 [hep-ph]} \BibitemShut
  {NoStop}%
\bibitem [{\citenamefont {Szafron}\ and\ \citenamefont
  {Czarnecki}(2016)}]{Szafron:2016cbv}%
  \BibitemOpen
  \bibfield  {author} {\bibinfo {author} {\bibfnamefont {R.}~\bibnamefont
  {Szafron}}\ and\ \bibinfo {author} {\bibfnamefont {A.}~\bibnamefont
  {Czarnecki}},\ }\bibfield  {title} {\bibinfo {title} {{Bound muon decay
  spectrum in the leading logarithmic accuracy}},\ }\href
  {https://doi.org/10.1103/PhysRevD.94.051301} {\bibfield  {journal} {\bibinfo
  {journal} {Phys. Rev.}\ }\textbf {\bibinfo {volume} {D94}},\ \bibinfo {pages}
  {051301} (\bibinfo {year} {2016})},\ \Eprint
  {https://arxiv.org/abs/1608.05447} {arXiv:1608.05447 [hep-ph]} \BibitemShut
  {NoStop}%
\bibitem [{\citenamefont {Heeck}\ \emph {et~al.}(2022)\citenamefont {Heeck},
  \citenamefont {Szafron},\ and\ \citenamefont {Uesaka}}]{Heeck:2021adh}%
  \BibitemOpen
  \bibfield  {author} {\bibinfo {author} {\bibfnamefont {J.}~\bibnamefont
  {Heeck}}, \bibinfo {author} {\bibfnamefont {R.}~\bibnamefont {Szafron}},\
  and\ \bibinfo {author} {\bibfnamefont {Y.}~\bibnamefont {Uesaka}},\
  }\bibfield  {title} {\bibinfo {title} {{Isotope dependence of muon decay in
  orbit}},\ }\href {https://doi.org/10.1103/PhysRevD.105.053006} {\bibfield
  {journal} {\bibinfo  {journal} {Phys. Rev. D}\ }\textbf {\bibinfo {volume}
  {105}},\ \bibinfo {pages} {053006} (\bibinfo {year} {2022})},\ \Eprint
  {https://arxiv.org/abs/2110.14667} {arXiv:2110.14667 [hep-ph]} \BibitemShut
  {NoStop}%
\bibitem [{\citenamefont {Kaygorodov}\ \emph {et~al.}(2025)\citenamefont
  {Kaygorodov}, \citenamefont {Kozhedub}, \citenamefont {Malyshev},
  \citenamefont {Davydov}, \citenamefont {Wu},\ and\ \citenamefont
  {Zhang}}]{Kaygorodov:2025yag}%
  \BibitemOpen
  \bibfield  {author} {\bibinfo {author} {\bibfnamefont {M.~Y.}\ \bibnamefont
  {Kaygorodov}}, \bibinfo {author} {\bibfnamefont {Y.~S.}\ \bibnamefont
  {Kozhedub}}, \bibinfo {author} {\bibfnamefont {A.~V.}\ \bibnamefont
  {Malyshev}}, \bibinfo {author} {\bibfnamefont {A.~O.}\ \bibnamefont
  {Davydov}}, \bibinfo {author} {\bibfnamefont {Y.}~\bibnamefont {Wu}},\ and\
  \bibinfo {author} {\bibfnamefont {S.~B.}\ \bibnamefont {Zhang}},\ }\bibfield
  {title} {\bibinfo {title} {{Study of atomic effects on electron spectrum in
  bound-muon decay process}}} (\bibinfo {year} {2025}),\ \bibinfo {note}
  {arXiv:2506.02416}\BibitemShut {NoStop}%
\bibitem [{\citenamefont {Fontes}\ and\ \citenamefont
  {Szafron}(2025{\natexlab{a}})}]{Fontes:2025xbt}%
  \BibitemOpen
  \bibfield  {author} {\bibinfo {author} {\bibfnamefont {D.}~\bibnamefont
  {Fontes}}\ and\ \bibinfo {author} {\bibfnamefont {R.}~\bibnamefont
  {Szafron}},\ }\bibfield  {title} {\bibinfo {title} {{QED corrections to
  bound-muon decays from an effective-field-theory framework}}} (\bibinfo
  {year} {2025}{\natexlab{a}}),\ \bibinfo {note} {arxiv:2510.26698}\BibitemShut
  {NoStop}%
\bibitem [{\citenamefont {Fontes}\ and\ \citenamefont
  {Szafron}(2025{\natexlab{b}})}]{Fontes:2025mps}%
  \BibitemOpen
  \bibfield  {author} {\bibinfo {author} {\bibfnamefont {D.}~\bibnamefont
  {Fontes}}\ and\ \bibinfo {author} {\bibfnamefont {R.}~\bibnamefont
  {Szafron}},\ }\bibfield  {title} {\bibinfo {title} {{EFT approach to the
  endpoint of muon decay-in-orbit}}} (\bibinfo {year} {2025}{\natexlab{b}}),\
  \bibinfo {note} {arxiv:2506.23021}\BibitemShut {NoStop}%
\bibitem [{\citenamefont {Fontes}\ and\ \citenamefont
  {Szafron}(2025{\natexlab{c}})}]{Fontes:2024yvw}%
  \BibitemOpen
  \bibfield  {author} {\bibinfo {author} {\bibfnamefont {D.}~\bibnamefont
  {Fontes}}\ and\ \bibinfo {author} {\bibfnamefont {R.}~\bibnamefont
  {Szafron}},\ }\bibfield  {title} {\bibinfo {title} {{An effective field
  theory for muon conversion and muon decay-in-orbit}},\ }\href
  {https://doi.org/10.1007/JHEP05(2025)171} {\bibfield  {journal} {\bibinfo
  {journal} {JHEP}\ }\textbf {\bibinfo {volume} {05}},\ \bibinfo {pages}
  {171}},\ \Eprint {https://arxiv.org/abs/2412.05702} {arXiv:2412.05702
  [hep-ph]} \BibitemShut {NoStop}%
\bibitem [{\citenamefont {\"Uberall}(1960)}]{Uberall:1960zz}%
  \BibitemOpen
  \bibfield  {author} {\bibinfo {author} {\bibfnamefont {H.}~\bibnamefont
  {\"Uberall}},\ }\bibfield  {title} {\bibinfo {title} {{Decay of $\mu^-$
  Mesons Bound in the K Shell of Light Nuclei}},\ }\href
  {https://doi.org/10.1103/PhysRev.119.365} {\bibfield  {journal} {\bibinfo
  {journal} {Phys. Rev.}\ }\textbf {\bibinfo {volume} {119}},\ \bibinfo {pages}
  {365} (\bibinfo {year} {1960})}\BibitemShut {NoStop}%
\bibitem [{\citenamefont {Webber}\ \emph {et~al.}(2011)\citenamefont {Webber}
  \emph {et~al.}}]{MuLan:2010shf}%
  \BibitemOpen
  \bibfield  {author} {\bibinfo {author} {\bibfnamefont {D.~M.}\ \bibnamefont
  {Webber}} \emph {et~al.} (\bibinfo {collaboration} {MuLan}),\ }\bibfield
  {title} {\bibinfo {title} {{Measurement of the Positive Muon Lifetime and
  Determination of the Fermi Constant to Part-per-Million Precision}},\ }\href
  {https://doi.org/10.1103/PhysRevLett.106.079901} {\bibfield  {journal}
  {\bibinfo  {journal} {Phys. Rev. Lett.}\ }\textbf {\bibinfo {volume} {106}},\
  \bibinfo {pages} {041803} (\bibinfo {year} {2011})},\ \Eprint
  {https://arxiv.org/abs/1010.0991} {arXiv:1010.0991 [hep-ex]} \BibitemShut
  {NoStop}%
\bibitem [{\citenamefont {Watanabe}\ \emph {et~al.}(1993)\citenamefont
  {Watanabe}, \citenamefont {Muto}, \citenamefont {Oda}, \citenamefont {Niwa},
  \citenamefont {Ohtsubo}, \citenamefont {Morita},\ and\ \citenamefont
  {Morita}}]{Watanabe:1993emp}%
  \BibitemOpen
  \bibfield  {author} {\bibinfo {author} {\bibfnamefont {R.}~\bibnamefont
  {Watanabe}}, \bibinfo {author} {\bibfnamefont {K.}~\bibnamefont {Muto}},
  \bibinfo {author} {\bibfnamefont {T.}~\bibnamefont {Oda}}, \bibinfo {author}
  {\bibfnamefont {T.}~\bibnamefont {Niwa}}, \bibinfo {author} {\bibfnamefont
  {H.}~\bibnamefont {Ohtsubo}}, \bibinfo {author} {\bibfnamefont
  {M.}~\bibnamefont {Morita}},\ and\ \bibinfo {author} {\bibfnamefont
  {R.}~\bibnamefont {Morita}},\ }\bibfield  {title} {\bibinfo {title}
  {Asymmetry and energy spectrum of electrons in bound-muon decay},\
  }\href@noop {} {\bibfield  {journal} {\bibinfo  {journal} {Atomic Data and
  Nucl. Data Tables}\ }\textbf {\bibinfo {volume} {54}},\ \bibinfo {pages}
  {165} (\bibinfo {year} {1993})}\BibitemShut {NoStop}%
\bibitem [{\citenamefont {Watanabe}\ \emph {et~al.}(1987)\citenamefont
  {Watanabe}, \citenamefont {Fukui}, \citenamefont {Ohtsubo},\ and\
  \citenamefont {Morita}}]{Watanabe:1987su}%
  \BibitemOpen
  \bibfield  {author} {\bibinfo {author} {\bibfnamefont {R.}~\bibnamefont
  {Watanabe}}, \bibinfo {author} {\bibfnamefont {M.}~\bibnamefont {Fukui}},
  \bibinfo {author} {\bibfnamefont {H.}~\bibnamefont {Ohtsubo}},\ and\ \bibinfo
  {author} {\bibfnamefont {M.}~\bibnamefont {Morita}},\ }\bibfield  {title}
  {\bibinfo {title} {{Angular distribution of electrons in bound muon decay}},\
  }\href@noop {} {\bibfield  {journal} {\bibinfo  {journal}
  {Prog.~Theor.~Phys.}\ }\textbf {\bibinfo {volume} {78}},\ \bibinfo {pages}
  {114} (\bibinfo {year} {1987})}\BibitemShut {NoStop}%
\bibitem [{\citenamefont {Aslam}\ \emph {et~al.}(2020)\citenamefont {Aslam},
  \citenamefont {Czarnecki}, \citenamefont {Zhang},\ and\ \citenamefont
  {Morozova}}]{Aslam:2020pqn}%
  \BibitemOpen
  \bibfield  {author} {\bibinfo {author} {\bibfnamefont {M.~J.}\ \bibnamefont
  {Aslam}}, \bibinfo {author} {\bibfnamefont {A.}~\bibnamefont {Czarnecki}},
  \bibinfo {author} {\bibfnamefont {G.}~\bibnamefont {Zhang}},\ and\ \bibinfo
  {author} {\bibfnamefont {A.}~\bibnamefont {Morozova}},\ }\bibfield  {title}
  {\bibinfo {title} {{Decay of a bound muon into a bound electron}},\
  }\href@noop {} {\bibfield  {journal} {\bibinfo  {journal} {Phys. Rev. D}\
  }\textbf {\bibinfo {volume} {102}},\ \bibinfo {pages} {073001} (\bibinfo
  {year} {2020})},\ \Eprint {https://arxiv.org/abs/2005.07276}
  {arXiv:2005.07276 [hep-ph]} \BibitemShut {NoStop}%
\bibitem [{\citenamefont {Varshalovich}\ \emph {et~al.}(1988)\citenamefont
  {Varshalovich}, \citenamefont {Moskalev},\ and\ \citenamefont
  {Khersonsky}}]{Varshalovich:1988ifq}%
  \BibitemOpen
  \bibfield  {author} {\bibinfo {author} {\bibfnamefont {D.~A.}\ \bibnamefont
  {Varshalovich}}, \bibinfo {author} {\bibfnamefont {A.~N.}\ \bibnamefont
  {Moskalev}},\ and\ \bibinfo {author} {\bibfnamefont {V.~K.}\ \bibnamefont
  {Khersonsky}},\ }\href@noop {} {\emph {\bibinfo {title} {{Quantum Theory of
  Angular Momentum: Irreducible Tensors, Spherical Harmonics, Vector Coupling
  Coefficients, 3nj Symbols}}}}\ (\bibinfo  {publisher} {World Scientific},\
  \bibinfo {address} {Singapore},\ \bibinfo {year} {1988})\BibitemShut
  {NoStop}%
\bibitem [{\citenamefont {Rose}(1961)}]{Rose:1961RET}%
  \BibitemOpen
  \bibfield  {author} {\bibinfo {author} {\bibfnamefont {M.~E.}\ \bibnamefont
  {Rose}},\ }\href@noop {} {\emph {\bibinfo {title} {Relativistic Electron
  Theory}}}\ (\bibinfo  {publisher} {John Wiley},\ \bibinfo {address} {New
  York},\ \bibinfo {year} {1961})\BibitemShut {NoStop}%
\bibitem [{\citenamefont {Salvat}\ \emph {et~al.}(1995)\citenamefont {Salvat},
  \citenamefont {Fern{\'a}ndez-Varea},\ and\ \citenamefont
  {Williamson~Jr}}]{salvat1995radial}%
  \BibitemOpen
  \bibfield  {author} {\bibinfo {author} {\bibfnamefont {F.}~\bibnamefont
  {Salvat}}, \bibinfo {author} {\bibfnamefont {J.~M.}\ \bibnamefont
  {Fern{\'a}ndez-Varea}},\ and\ \bibinfo {author} {\bibfnamefont
  {W.}~\bibnamefont {Williamson~Jr}},\ }\bibfield  {title} {\bibinfo {title}
  {{Accurate numerical solution of the radial Schr{\"o}dinger and Dirac wave
  equations}},\ }\href@noop {} {\bibfield  {journal} {\bibinfo  {journal}
  {Computer Physics Communications}\ }\textbf {\bibinfo {volume} {90}},\
  \bibinfo {pages} {151} (\bibinfo {year} {1995})}\BibitemShut {NoStop}%
\bibitem [{\citenamefont {Salvat}\ and\ \citenamefont
  {Fern{\'a}ndez-Varea}(2019)}]{salvat2019radial}%
  \BibitemOpen
  \bibfield  {author} {\bibinfo {author} {\bibfnamefont {F.}~\bibnamefont
  {Salvat}}\ and\ \bibinfo {author} {\bibfnamefont {J.~M.}\ \bibnamefont
  {Fern{\'a}ndez-Varea}},\ }\bibfield  {title} {\bibinfo {title} {{RADIAL: A
  Fortran subroutine package for the solution of the radial Schr{\"o}dinger and
  Dirac wave equations}},\ }\href@noop {} {\bibfield  {journal} {\bibinfo
  {journal} {Computer Physics Communications}\ }\textbf {\bibinfo {volume}
  {240}},\ \bibinfo {pages} {165} (\bibinfo {year} {2019})}\BibitemShut
  {NoStop}%
\bibitem [{\citenamefont {Piessens}\ \emph {et~al.}(2012)\citenamefont
  {Piessens}, \citenamefont {de~Doncker-Kapenga}, \citenamefont
  {{\"U}berhuber},\ and\ \citenamefont {Kahaner}}]{piessens2012quadpack}%
  \BibitemOpen
  \bibfield  {author} {\bibinfo {author} {\bibfnamefont {R.}~\bibnamefont
  {Piessens}}, \bibinfo {author} {\bibfnamefont {E.}~\bibnamefont
  {de~Doncker-Kapenga}}, \bibinfo {author} {\bibfnamefont {C.~W.}\ \bibnamefont
  {{\"U}berhuber}},\ and\ \bibinfo {author} {\bibfnamefont {D.~K.}\
  \bibnamefont {Kahaner}},\ }\href@noop {} {\emph {\bibinfo {title} {{QUADPACK:
  a subroutine package for automatic integration}}}}\ (\bibinfo  {publisher}
  {Springer},\ \bibinfo {address} {Berlin},\ \bibinfo {year}
  {2012})\BibitemShut {NoStop}%
\end{thebibliography}
%

\end{document}